\documentclass[aps,pre,twocolumn,amsmath,amssymb,superscriptaddress,floatfix,showpacs]{revtex4}
\usepackage{amssymb,hyperref}
\usepackage{graphicx}
\usepackage{subfigure}

\begin{document}

\title{Saddles, Arrows, and Spirals: Deterministic Trajectories in Cyclic
Competition of Four Species}
\author{C. H. Durney, S. O. Case, M. Pleimling, and R.K.P. Zia}
\affiliation{Department of Physics, Virginia Tech,\\
Blacksburg, Virginia, 24061-0435, USA}
\date{\today }

\begin{abstract}
Population dynamics in systems composed of cyclically competing species has
been of increasing interest recently. Here, we investigate a system with
four or more species. Using mean field theory, we study in detail the
trajectories in configuration space of the population fractions. We discover
a variety of orbits, shaped like saddles, spirals, and straight lines. Many
of their properties are found explicitly. Most remarkably, we identify a
collective variable which evolves simply as an exponential: $\mathcal{Q}%
\propto e^{\lambda t}$, where $\lambda$ is a function of the reaction rates. 
It provides information on the state of the system for late
times (as well as for $t\rightarrow -\infty $). We discuss implications of
these results for the evolution of a finite, stochastic system. A
generalization to an arbitrary number of cyclically competing species yields
valuable insights into universal properties of such systems.
\end{abstract}
\maketitle

%\input epsf.sty

%\eads{\mailto{Michel.Pleimling@vt.edu}, \mailto{rkpzia@vt.edu}}

\section{Introduction}

In the study of population dynamics for biological systems or chemical
reactions, it is customary to start with a set of ordinary
differential/difference equations which model the time evolution of
populations or concentrations. These would be appropriate if the various
constituents of the system are well mixed (so that spatial structures can be
ignored){\em \ and} if the stochastic aspects of the evolution can be
overlooked. In this sense, the term `mean field theory' (MFT) is most
appropriate, as both spatial and temporal fluctuations are absent. Despite
these limitations, MFT has proven valuable in several respects. As in other
areas of statistical physics, it typically provides an adequate description
in much of the space of (control) parameters, e.g., Landau's theory of phase
transitions. In addition, it yields significant insights into many
interesting phenomena associated with the underlying non-linear dynamics. A
well known example is the simple logistic map, which gives rise to a rich
structure of bifurcation, chaos, and universality \cite{May76,Feig78,Feig79}%
. Finally, it provides a stage on which hidden symmetries can be showcased
readily.

In this context, we are motivated to consider MFT for a simple model of
population dynamics, involving four species competing cyclically. This is a
generalization of systems with three cyclically competing species, also
known as the rock-paper-scissors games, which attracted considerable
attention in recent years \cite
{Fra96a,Fra96b,Pro99,Tse01,Ker02,Kir04,Rei06,Rei07,Rei08,Cla08,Pel08,Rei08a,Ber09,Ven10,Shi10,And10,Rul10,Wan10,Mob10,He10,Win10,Dob10}%
. Labeling the species $a,b,c,d$, we allow $a$ to `prey on' $b$ with some
rate, and $b,c,d$ to `prey on' $c,d,a$, respectively, with some other rates.
Unlike earlier studies, we will not restrict our rates, but consider them in
general. As we will show, apart from having a much larger and complex
parameter space, the four species system (4SS) is qualitatively different.
Indeed, in MFT, the evolution of the three species system (3SS) is
relatively trivial, since a non-linear invariant (a hidden symmetry) renders
all trajectories into closed orbits on a plane. As a result, there is little
hint of the presence of absorbing states, which are of fundamental interest
for the study of extinction probabilities in a stochastic system. By
contrast, such a scenario is not the case, typically, for the 4SS. In
addition, when one species becomes extinct, our problem reduces to an
apparently trivial limit of the 3SS. Yet, the MFT offers useful and
quantitative predictions. In a separate publication \cite{CDPZ-JSTAT}, we
will present extensive simulations of the full stochastic model and show
that many of the MFT predictions are well born out. In this article, we will
focus only on the deterministic MFT and the rich variety of phenomena it
predicts. Though some of the preliminary results on the 4SS have been
presented earlier \cite{CDPZ-epl}, this paper will delve into not only
details of the analysis, but also provide new insights into the behavior of
systems with $S>4$ species.

Before we begin, we should mention that there are studies for the general $S$
case, though far few in number than for the 3SS. Most of these studies
considered systems on one- or two-dimensional lattices \cite
{Fra96a,Fra96b,Fra98,Kob97,Sat02,Sza04,He05,Sza07b,Sza08,Dob10}. Due to the
presence of spatial structure, these systems display far more complex
behavior and so, the authors typically restrict themselves to special sets
of rates (e.g., all equal, or all but one being equal). Finally, some
studies, motivated by real-world systems, focused on a large number of
competing species with more complicated interactions schemes \cite
{Sil92,Sza01}. Here, we will focus on phenomena that emerge from arbitrary
(positive) rates, but consider only systems with {\em no spatial structure}.
An interesting application of a four-species model without spatial structure
to the endogenous and exogenous origins of diseases can be found in \cite{Sor09}.

This paper is organized as follows. Though our main focus will be MFT, we
will first discuss how the differential equations arise from a fully
stochastic, microscopic model of the 4SS. Details of the latter (which can
be simulated by Monte Carlo methods) and the associated master equation will
be presented in Section II. The approximation scheme which leads
to the MFT will also be provided. Section III will be devoted to the
analysis of the MFT equations, with a number of explicit results. Their
implications for the stochastic evolution will be considered in Section IV.
By generalizing to systems with $S>4$ cyclically competing species, we gain
deeper insights into how certain aspects of the solutions are `universal.'
These considerations will be presented in Section V. Readers who are
comfortable with abstract formulations will find that some of the
conclusions in Sections III are just special cases. Finally, we will
summarize our findings in a concluding section and sketch possible avenues
for future research.

Interestingly, we found \cite{CDPZ-epl} that, unlike in the 3SS, the four
species form partner-pairs, much like in the game of Bridge. In a system
with $N$ individuals, there are $2\left( N+1\right) $ absorbing states, most
of which consist of a surviving partner-pair. Since MFT is best suited for
systems with large $N$, we should expect that all the final states in this
approximation to consists of a coexisting pair: either $a$-$c$ or $b$-$d$.
Of course, many interesting issues associated with the full, stochastic
model (e.g., extinction probabilities) cannot be answered by MFT, while
simulations reveal complex extinction scenarios that depend on both the
predation rates and initial conditions. Since those investigations are quite
extensive, they will be reported elsewhere \cite{CDPZ-JSTAT}.

\section{The microscopic stochastic model and its mean field approximation}

Our four species system consists of $N_m$ individuals of species $m=\left(
a,b,c,d\right) $. With no spatial structure, any individual can interact
with any other, in a cyclically competing manner. In a time step of a fully
stochastic model, a random pair is chosen and, if they are `{\em cyclically}
different,' the following interactions occur with probabilities $p_m:$ 
\begin{eqnarray*}
&&a+b\stackrel{p_a}{\rightarrow }a+a;\,\,b+c\stackrel{p_b}{\rightarrow }b+b
\\
&&c+d\stackrel{p_c}{\rightarrow }c+c;\,\,d+a\stackrel{p_d}{\rightarrow }d+d
\end{eqnarray*}
Thus, the total number in the system 
\[
N=\sum_mN_m 
\]
is a constant and the full configuration space is just a set of points
within a regular tetrahedron (of length $N$ on each side, with the four
vertices being $N=N_m$ for one of the $m$'s). Of course, we will later
consider the fractions $N_m/N$, which are natural variables in the large $N$
limit and the mean field approximation.

Let us emphasize that $ac$ and $bd$ pairs do {\em not }interact, so that it
is possible for the final (absorbing) state to display coexistence of these
pairs. In other words, any point along the $a$-$c$ and $b$-$d$ edges of the
tetrahedron represents such a state. Indeed, this property provides the
first major difference between our 4SS and the simpler 3SS: We have $2\left(
N+1\right) $ absorbing states here instead of just $3$ (regardless of $N$).
We should also remark that each face of the tetrahedron is also `absorbing'
in the sense that transitions into the face are irreversible. Within each
face, the problem is a special limit of the 3SS, namely, one of the three
rates being zero.

Given the interaction rules, the master equation for $P\left( \left\{
N_m\right\} ;\tau \right) $, the probability for finding the system $\tau $
steps after a particular initial distribution $P_0\left( \left\{ N_m\right\}
\right) $, can be easily written. It provides the change in $P$ over one
step: 
\begin{eqnarray}
&&P\left( \left\{ N_m\right\} ;\tau +1\right) -P\left( \left\{ N_m\right\}
;\tau \right)  \nonumber \\
&=&\frac{p_a\left( N_a-1\right) \left( N_b+1\right) }{N\left( N-1\right) /2}%
P\left( N_a-1,N_b+1,N_c,N_d;\tau \right)  \nonumber \\
&&+\frac{p_b\left( N_b-1\right) \left( N_c+1\right) }{N\left( N-1\right) /2}%
P\left( N_a,N_b-1,N_c+1,N_d;\tau \right)  \nonumber \\
&&+\frac{p_c\left( N_c-1\right) \left( N_d+1\right) }{N\left( N-1\right) /2}%
P\left( N_a,N_b,N_c-1,N_d+1;\tau \right)  \nonumber \\
&&+\frac{p_d\left( N_d-1\right) \left( N_a+1\right) }{N\left( N-1\right) /2}%
P\left( N_a+1,N_b,N_c,N_d-1;\tau \right)  \nonumber \\
&&-\frac Z{N\left( N-1\right) /2}P\left( \left\{ N_m\right\} ;\tau \right)
\label{PRP-ME}
\end{eqnarray}
where 
\begin{equation}
Z\equiv p_aN_aN_b+p_bN_bN_c+p_cN_cN_d+p_dN_dN_a\,\,.  \label{Z-def}
\end{equation}
For the initial $P_0\left( \left\{ N_m\right\} \right) $, it is sufficient
to use $\delta $ distributions. Indeed, every simulation run begins with a
single point (e.g., the symmetry point $N_m=N/4$)! Since the master equation
is linear, the evolution starting with any other distribution is just the
sum of the $P$'s that begin as $\delta $'s at various appropriate points.

{}From the master equation for $P\left( \left\{ N_m\right\} ;\tau \right) $,
we can easily derive a partial differential equation for the associated
generating function. But, to find the solution for either equation is far
from trivial. Instead, we turn to the mean field approximation, which should
be adequate for large $\left\{ N_m\right\} $. Thus, we can expect break
downs near the extinction of one or more species, i.e., near an `absorbing
face' or an absorbing state.

In a MFT, the focus is the evolution of the averages of say, the fractions 
$N_m/N$, denoted by 
\begin{eqnarray*}
A\left( \tau \right) &\equiv &\left\langle N_a/N\right\rangle _\tau \equiv
\sum_{\left\{ N_m\right\} }\left( N_a/N\right) P\left( \left\{ N_m\right\}
;\tau \right) \,\,, \\
B\left( \tau \right) &\equiv &\left\langle N_b/N\right\rangle _\tau ,
\,C\left( \tau \right) \equiv \left\langle N_c/N\right\rangle _\tau ,
\,D\left( \tau \right) \equiv \left\langle N_d/N\right\rangle _\tau
\end{eqnarray*}
Of course, we have the constraint 
\begin{equation}
A+B+C+D=1  \label{sum}
\end{equation}
so that there are only three independent variables.

Following standard routes, we can derive an equation for the changes, 
$A\left( \tau +1\right) -A\left( \tau \right) $, etc., from the master
equation (\ref{PRP-ME}). These will involve averages of products (e.g., 
$\left\langle N_aN_b\right\rangle _\tau $) on the right. The mean field
approximation consists of neglecting all correlations and replacing the
averages of products by the products of averages, so that the end result is
a closed set of deterministic equations for $A\left( \tau \right) $, 
$B\left( \tau \right) $, etc. Finally, by taking the $N\rightarrow \infty $
limit, defining $t\equiv \tau /N$ (which can be regarded as continuous), and
letting $A\left( \tau +1\right) -A\left( \tau \right) \rightarrow \left(
1/N\right) \partial _tA\left( t\right) $, we arrive at the MFT equations:

\begin{eqnarray}
\partial _tA &=&\left[ k_aB-k_dD\right] A  \label{ABCD eqns 1} \\
\partial _tB &=&\left[ k_bC-k_aA\right] B  \label{ABCD eqns 2} \\
\partial _tC &=&\left[ k_cD-k_bB\right] C  \label{ABCD eqns 3} \\
\partial _tD &=&\left[ k_dA-k_cC\right] D  \label{ABCD eqns 4}
\end{eqnarray}
In these time units, the `rates' $k_m$ can be related to the original
probabilities via $k_m=2p_m$. In the literature, however, time is often
rescaled again, so as to normalize the rates to 
\begin{equation}
k_a+k_b+k_c+k_d=1  \label{k-norm}
\end{equation}
as well.

\section{Analysis of MFT equations}

This section will be devoted to the implications of eqns. (\ref{ABCD eqns 1}-%
\ref{ABCD eqns 4}), assuming the initial fractions are $\left(
A_0,B_0,C_0,D_0\right) $. Given the normalization (\ref{k-norm}), our
parameter space is, in general, quite large: 6-dimensional. We will begin
with a restricted problem -- evolution on one of the faces of the
tetrahedron -- which is explicitly solvable.

\subsection{Evolution when one species is extinct}

\begin{figure}
\vglue 5mm
\begin{center}
\includegraphics[width=8.cm,angle=0]{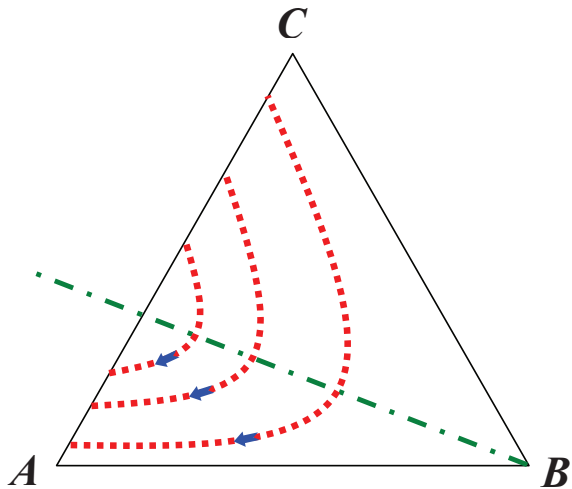}
\end{center}
\caption{(Color online) Schematic plot of trajectories in the $(A, B, C)$ plane
in case species $d$ dies first. The dashed red lines show the generalized
hyperbolas (\ref{R-def}) along which the system evolves by increasing A. At the
intersections of the hyperbolas and the dot-dashed green lines $B$ is maximal.
}
\label{fig1}
\end{figure}

Clearly, this restricted problem is not only much simpler than the full 4SS,
it is also considerably simpler than the general 3SS. The reason is obvious:
One of the three remaining species does not `consume' anyone and so, its
numbers can never increase. Without loss of generality, let us consider the
face with $D=0$ with arbitrary initial $\left( A_0,B_0,C_0\right) $. In this
case, neither $k_c$ nor $k_d$ plays a role, while eqns. (\ref{ABCD eqns 1}-%
\ref{ABCD eqns 4}) reduce significantly. As a result, 
\begin{equation}
R\equiv A^{k_b}C^{k_a}  \label{R-def}
\end{equation}
is an invariant (similar to that in the full 3SS \cite{Ber09}), so that it is
always fixed at $A_0^{k_b}C_0^{k_a}$. Of course, our problem now reduces to
one with a single variable, which we choose to be 
\begin{equation}
V\equiv A^{k_b}C^{-k_a}  \label{S-def}
\end{equation}
so that its evolution is monotonically increasing until $B$ vanishes: 
\begin{eqnarray}
\partial _t\ln V=2k_ak_bB\geq 0  \label{S-eqn}
\end{eqnarray}
Meanwhile, $B=1-A-C$ and $A^{2k_b}$ ($C^{2k_a}$) is given by $RV$ ($R/V$),
so that the full solution for $V\left( t\right) $ can be obtained through
the integral 
\begin{eqnarray}
\int_{V_0}^{V\left( t\right) }\frac{dv}{v\left[ 1-\left( Rv\right)
^{1/2k_b}-\left( R/v\right) ^{1/2k_a}\right] }=2k_ak_bt  \label{S-sol}
\end{eqnarray}
where $V_0\equiv A_0^{k_b}C_0^{-k_a}$. Though analytically precise, this
form of the solution does not provide much insight on the evolution of the
system. Instead, a schematic plot of sample trajectories, as shown in Fig. \ref{fig1},
should be much more helpful. Representing various
values of $R$, the curved lines (dashed, red online) are generalized
hyperbolas (\ref{R-def}). The initial fractions $\left( A_0,B_0,C_0\right) $
dictate which curve that system will follow, while the evolution takes it
towards larger $A$ (in the direction of the arrows, blue online), ending at
a point $\left( A_f,0,C_f\right) $ on the $a$-$c$ line. Note that the
straight line (dot-dashed, green online), given by $k_bC=k_aA$, intersects
each of these curves at right angles, corresponding to $B$ being maximal at
these points. Finally, by exploiting the invariant $R$, the values $A_f$ and 
$C_f$ ($=1-A_f$) can be found from a (generally transcendental) equation,
e.g., 
\[
A_f^{k_b}\left( 1-A_f\right) ^{k_a}=A_0^{k_b}C_0^{k_a}\,\,.
\]
Since $k_a,k_b\in \left( 0,1\right) $, a schematic plot of the left hand
side reveals that there are two solutions. Of these, the larger one is the
final $A_f$, since $\partial _tA\geq 0$. Of course, numerical methods will
be needed typically for obtaining the explicit $A_f$. These considerations
are useful when MFT predictions are compared to simulation data for
times after one of the four species goes extinct \cite{CDPZ-epl, CDPZ-JSTAT}.

%%%%%%%%%%% FIG 1  %%%%%%%%%%%%%%%%%%%%%%%%%%%%%%%
%%%%%%%%%%%%%%%%%%%%%%%%%%%%%%%%%%%%%%%%%%%%%%

\subsection{A general criterion for survival/extinction}

Turning to the full 4SS, we identify a single parameter that determines
which of the pairs survive in the long time limit. To find this criterion,
we note that eqns. (\ref{ABCD eqns 1}-\ref{ABCD eqns 4}) can be written as 
\begin{eqnarray}
\partial _t\ln A &=&k_aB-k_dD;\quad \partial _t\ln C=k_cD-k_bB  \label{lnAC}
\\
\partial _t\ln B &=&k_bC-k_aA;\quad \partial _t\ln D=k_dA-k_cC  \label{lnBD}
\end{eqnarray}
since none of the fractions vanishes in finite $t$. This form is quite
natural, since exponential growth/decay in populations is common. It also
exposes clearly the coupling between the pairs $a$-$c$ and $b$-$d$.
Constructing appropriate linear combinations, we showcase the contributions
from a {\em single} species to the growth/decay of two other species: 
\begin{eqnarray*}
\partial _t\left[ k_b\ln A+k_a\ln C\right] &=&\lambda D \\
\partial _t\left[ k_c\ln A+k_d\ln C\right] &=&\lambda B \\
\partial _t\left[ k_c\ln B+k_b\ln D\right] &=&-\lambda A \\
\partial _t\left[ k_d\ln B+k_a\ln D\right] &=&-\lambda C
\end{eqnarray*}
where we have introduced the key control parameter 
\begin{equation}
\lambda \equiv k_ak_c-k_bk_d\,\,.  \label{lambda-def}
\end{equation}
Adding and subtracting appropriately, we see that the quantity 
\begin{equation}
Q\equiv \frac{A^{k_b+k_c}C^{k_d+k_a}}{B^{k_c+k_d}D^{k_a+k_b}}  \label{Q-def}
\end{equation}
evolves in an extremely simple manner: 
\begin{equation}
Q\left( t\right) =Q\left( 0\right) e^{\lambda t}~.  \label{Q-evol}
\end{equation}
The form of $Q$ also points to a simple interpretation of the behavior of
the species. As in the card game, Bridge, $a$-$c$ and $b$-$d$ form {\em %
opposing partner-pairs}. For example, $a$'s action (consuming $b$) benefits $%
c$, while $c$'s action benefits $a$. Thus, we may refer to $a$-$c$ and $b$-$%
d $ as partners. Since no fraction can exceed unity, the only way for $Q$ to
vanish or diverge is for one of the pairs to go extinct. Thus, the sign of $%
\lambda $ is the selection criterion for, and $Q$ is a quantitative measure
of, which pair survives.

In this connection, we see that, unlike the 3SS \cite{Ber09}, the `weakest'
is not necessarily the survivor (or among the survivors) here. Indeed, if
both members of a partner pair are weak (compared to their opponents), then
the sign of $\lambda $ predicts the eventual demise of this pair, while the
magnitude of $\lambda $ hints at how rapidly they will die out. As we have
pointed out \cite{CDPZ-epl}, the general maxim seems to be: ``The prey of
the prey of the weakest is the {\em least likely} to survive.'' In the 3SS,
this maxim also applies and is consistent with `survival of the weakest.'
There, `the prey of the prey of the weakest' happens to be its predator, the
demise of which naturally enhances its survival.

\subsection{Saddle shaped orbits and fixed points for systems with $%
k_ak_c=k_bk_d$}

Clearly, special properties will appear in a 4SS with $\lambda =0$, when $Q$
becomes an invariant. Indeed, there are {\em two} invariants \cite
{Sza07,Dob10,CDPZ-epl}. Neither pair goes extinct and the system evolves
along periodic, closed orbits in configuration space. It is tempting to
regard $Q$ as the generalization of the quantity $R\equiv
A^{k_b}B^{k_c}C^{k_a}$ (introduced in \cite{Ber09}) in the 3SS, except that $%
R$ is invariant for {\em any} set of rates. As will be shown in Section V,
there are also fundamental differences between systems with even and odd
number of cyclically competing species. As a result, direct comparisons
between $R$ and $Q$ are not helpful, though both play important roles in
identifying collective degrees of freedom that evolve trivially.

\begin{figure}[h]
\vglue 5mm
\begin{center}
\includegraphics[width=7.cm,angle=0]{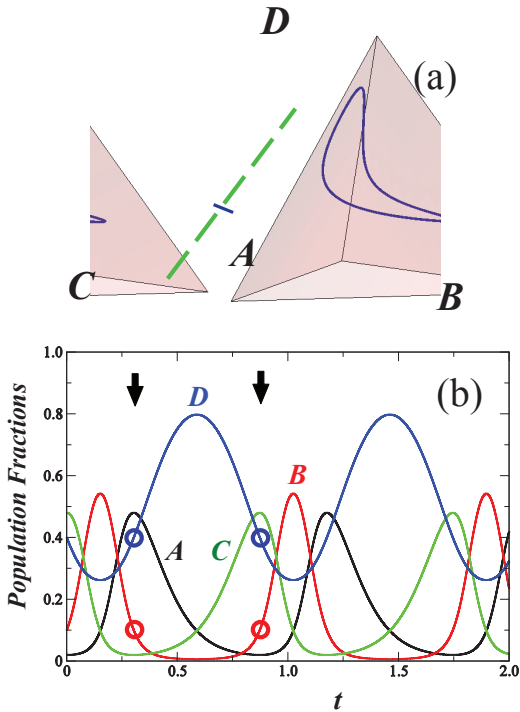}
\end{center}
\caption{(Color online) 
(a) Example of a closed loop (solid curve) in the tetrahedron forming 
the configuration space, encircling the line of fixed points (green 
dashed line). Note that the fixed line bridges absorbing states on the 
$a$-$c$ and $b$-$d$ edges.  The loop is generated by solving the 
mean field equations with a fourth order Runge-Kutta scheme, using 
time step $t=0.00001$. The rates $( k_a , k_b , k_c , k_d )$ used are
$(0.4, 0.4, 0.1, 0.1)$, yielding $\lambda =0$. 
(b) Evolution of the four species fractions as a function of time. 
The initial fractions $( A_0 , B_0 , C_0 , D_0 )$ are 
$( 0.02, 0.10, 0.48, 0.40 )$. 
Next to each curve is its species label. The red and blue open 
circles mark the values of $B$ and $D$, respectively, associated with 
the two turning points of $A$, located by the black arrows. Note that
both red (blue) circled values are the same. See text for details.
}
\label{fig2}
\end{figure}

Proceeding to investigate the extra invariant besides $Q$, we find an
appealing, symmetric way to display the two constants of motion: 
\begin{equation}
f\equiv A^{k_b}C^{k_a}\quad \mbox{and}\quad g\equiv B^{k_d}D^{k_a}
\label{f,g}
\end{equation}
(or equivalently, $f^{\prime }\equiv A^{k_c}C^{k_d}\quad $and$\quad
g^{\prime }\equiv B^{k_c}D^{k_b}$). These generalize the products $AC$ and $%
BD$ in \cite{Sza07}, where the rates are all unity. Since invariants are
fixed by the initial conditions ($f=A_0^{k_b}C_0^{k_a}$, etc.), it is
natural to define the variables 
\begin{eqnarray}
\rho _a &\equiv &\left( A/A_0\right) ^{k_b};\quad \rho _c\equiv \left(
C/C_0\right) ^{k_a}  \label{rho-def1} \\
\rho _b &\equiv &\left( B/B_0\right) ^{k_d};\quad \rho _d\equiv \left(
D/D_0\right) ^{k_a}  \label{rho-def2}
\end{eqnarray}
Then, the constants of motion are simply the products: 
\begin{equation}
\rho _a\rho _c=1;\quad \rho _b\rho _d=1~.  \label{rho-eqns}
\end{equation}
They define hyperbolic sheets through the tetrahedron and their intersection
is a closed loop that resembles (the rim of) a saddle. Fig. \ref{fig2}a
shows an example of such an orbit, for the case $k_a=0.4$, $k_b=0.4$, 
$k_c=0.1$, $k_d=0.1$. Here, the MFT eqns. 
(\ref{ABCD eqns 1}-\ref{ABCD eqns 4}) have been integrated numerically 
using a standard fourth-order Runge-Kutta scheme. 
Fig. \ref{fig2}b shows the ever-lasting oscillations in $A,...,D$
associated with this orbit.

%%%%%%%%%%%%%%%%%%%%%%%%%%%%%%%%%%%%%%%Fig. 2%%%%%%%%%%%%%%%%%%%%%%%%%%%%%
%%%%%%%%%%%%%%%%%%%%%%%%%%%%%%%%%%%%%%%Fig. 2%%%%%%%%%%%%%%%%%%%%%%%%%%%%%

Many characteristics of a closed orbit in our tetrahedron are accessible
analytically. We will summarize the findings here and defer the details to
the Appendix. Let us start with the simplest: the average position (over one
period, $T$): 
\[
\stackrel{\circ }{A}\equiv \frac 1T\int_0^TA\left( t\right) dt;\quad 
\stackrel{\circ }{B}\equiv \frac 1T\int_0^TB\left( t\right) dt;\quad etc. 
\]
Integrating eqns. (\ref{lnAC},\ref{lnBD}), we find that the left vanish
identically. Thus, we conclude 
\begin{equation}
k_a\stackrel{\circ }{A}=k_b\stackrel{\circ }{C};\quad k_a\stackrel{\circ }{B}%
=k_d\stackrel{\circ }{D}\,\,.  \label{AB-circle}
\end{equation}
Below, we will discuss the presence of a line of fixed-points for $\lambda
=0 $ systems, its properties given by the rates solely. It is perhaps not
surprising that $\left( \stackrel{\circ }{A},\stackrel{\circ }{B},\stackrel{%
\circ }{C},\stackrel{\circ }{D}\right) $ lies on this line. Its precise
location depends on the initial conditions, but finding this explicit
dependence remains elusive. Another conclusion from this consideration is
that all closed orbits enclose this non-trivial fixed line.

A more detailed description of our closed orbit is its projection onto, say,
the $\rho _a$-$\rho _d$ plane. One possible representation is to exploit
eqns. (\ref{rho-def1},\ref{rho-def2},\ref{sum}): 
\begin{equation}
A_0\rho _a^{1/k_b}+B_0\rho _d^{-1/k_d}+C_0\rho _a^{-1/k_a}+D_0\rho
_d^{1/k_a}=1\,\,.  \label{saddle1}
\end{equation}
In the Appendix, we will show how this ungainly looking equation can be
transformed into the familiar equation for a circle: $\alpha ^2+\beta ^2=1$.
We can gain some insight into this orbit by studying the extremal points. In
particular, focusing on $A$ and $D$ of an orbit, we denote the extremes
through the expressions:

\[
A\in \left[ A_{-},A_{+}\right] ,\qquad D\in \left[ D_{-},D_{+}\right] \,\,.
\]
These extremal points are given by the solution to the following equations: 
\begin{eqnarray}
A_{\pm }+J_AA_{\pm }^{-k_b/k_a} &=&1-K_A\left( B_0^{k_d}D_0^{k_a}\right)
^{1/\left( k_a+k_d\right) }  \label{A-hat eqn} \\
D_{\pm }+J_DD_{\pm }^{-k_a/k_d} &=&1-K_D\left( A_0^{k_b}C_0^{k_a}\right)
^{1/\left( k_a+k_b\right) }\,\,.  \label{D-hat eqn}
\end{eqnarray}
where $J$ and $K$ are constants: 
\begin{eqnarray}
J_A &\equiv &C_0A_0^{k_b/k_a},\quad J_D\equiv B_0D_0^{k_a/k_d} \\
K_A &\equiv &\left( \frac{k_a}{k_d}\right) ^{k_d/\left( k_a+k_d\right)
}+\left( \frac{k_d}{k_a}\right) ^{k_a/\left( k_a+k_d\right) }  \label{KA} \\
K_D &\equiv &\left( \frac{k_b}{k_a}\right) ^{k_a/\left( k_a+k_b\right)
}+\left( \frac{k_a}{k_b}\right) ^{k_b/\left( k_a+k_b\right) }\,\,.
\label{KD}
\end{eqnarray}
Being transcendental in general, these can be solved only numerically.
Nevertheless, we can appreciate that there are typically two solutions to
each (by a simple sketch of, say, $x+x^{-p}$). Obviously, the smaller
(larger) of the two is the minimum (maximum). Interestingly, when a species
is at its extremum, the fractions of its opposing partner-pair assume the 
{\em same value}. For example, when $A$ is at {\em either} turning point ($%
A_{\pm }$), $B$ takes on the same value, $\left[
k_d^{k_a}k_a^{-k_a}B_0^{k_d}D_0^{k_a}\right] ^{1/\left( k_a+k_d\right) }$.
Similarly, $D$ takes on one value here: $\left[
k_a^{k_d}k_d^{-k_d}B_0^{k_d}D_0^{k_a}\right] ^{1/\left( k_a+k_d\right) }$.
To illustrate, we mark one set in Fig. \ref{fig2}b with open circles.
As will be discussed in Section IV, the minima $\left(
A_{-},B_{-},C_{-},D_{-}\right) $ can play a role in predicting survivability
in stochastic systems.

Finally, like the 3SS, there are non-trivial fixed points. Indeed, with one
more degree of freedom than 3SS, we find a {\em line} of such points here.
Note that these are neutrally stable, unlike the trivially time-independent,
absorbing states. The simplest route to this line is to set the right of
eqns. (\ref{ABCD eqns 1}-\ref{ABCD eqns 4}) to zero. Denoting the values on
this line by $\left( A^{*},B^{*},C^{*},D^{*}\right) $, we see that they
satisfy $k_aA^{*}=k_bC^{*}$ and $k_aB^{*}=k_dD^{*}$. In other words, the
line is the intersection of the two planes. Exploiting $A+B+C+D=1$, we find
an elegant way to display these values. Introducing a parameter $\gamma \in
\left[ 0,1\right] $, we have 
\begin{equation}
\left( A^{*},C^{*}\right) =\frac{\left( k_b,k_a\right) }{k_a+k_b}\gamma
;\quad \left( B^{*},D^{*}\right) =\frac{\left( k_d,k_a\right) }{k_a+k_d}%
\left( 1-\gamma \right)\, .   \label{FL}
\end{equation}
Clearly, this line is straight, and bridges the two absorbing states on the $%
a$-$c$ and $b$-$d$ lines, as illustrated in Fig. \ref{fig2}a.
For the neighborhood of this
line, it is natural to study linearized versions of eqns. (\ref{ABCD eqns 1}-%
\ref{ABCD eqns 4}). Writing $A=A^{*}+\Delta _a$ with $\Delta _a\ll A^{*}$, $%
B=B^{*}+\Delta _b$ with $\Delta _b\ll B^{*}$, etc., we first note that $%
\sum_m\Delta _m=0$ and $\partial _t\left( \Delta _a+\Delta _c\right)
=0=\partial _t\left( \Delta _b+\Delta _d\right) $. Then, by defining $%
q\equiv k_a\Delta _a-k_b\Delta _c$ and $p\equiv k_a\Delta _b-k_d\Delta _d$,
we can cast the linearized equations in the form of a standard simple
harmonic oscillator: $\dot{q}=\left( k_a+k_b\right) A^{*}p;\,\,\dot{p}%
=-\left( k_a+k_d\right) B^{*}q$. In other words, the saddle-like orbits
become planar while the system evolves along ellipses, with angular
frequency $\sqrt{k_bk_d\gamma \left( 1-\gamma \right) }$. Remarkably,
contrary to natural expectations, the fixed line is not normal to the orbit
plane in general.

To summarize, if $\lambda =0$, the evolution is relatively simple. Every
starting point is associated with a closed, saddle-shaped orbit, on which
the system remains forever. In general, analytic closed form expressions for
them are not known (as far as we are aware). However, all enclose a fixed
line and many of their properties can be computed analytically. Obviously,
for finite systems evolving stochastically, these conclusions are valid only
approximately -- as long as the system is far from absorbing states.

\subsection{Systems with $k_ak_c\neq k_bk_d$ : spirals and arrows}

With such rates, $Q$ grows/decays exponentially, so that non-trivial fixed
points cannot exist. Since all fractions are bounded by unity, this implies
that either $BD$ or $AC$ vanishes in the large $t$ limit. In other words,
for a system with finite $N$, extinction of one of the species must occur
quite rapidly, especially if $\lambda $ is large. Before we continue to the
analysis of these cases, let us provide a simple and intuitive picture, for
appreciating the significance of $\lambda $.

%%%%%%%%%%% FIG 3  %%%%%%%%%%%%%%%%%%%%%%%%%%%%%%%
\begin{figure}
\vglue 5mm
\begin{center}
\includegraphics[width=8.cm,angle=0]{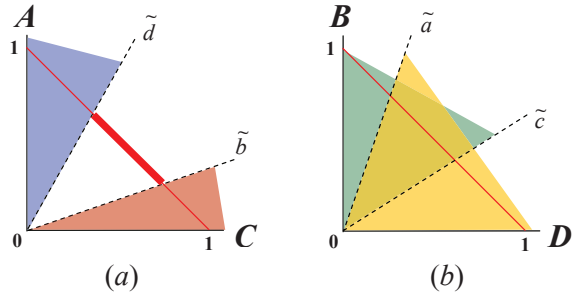}
\end{center}
\caption{(Color online) (a) $A$-$C$ and (b) $B$-$D$ plane. When a trajectory
enters the unshaded region in (a) and the doubly shaded region in (b), both 
$B$ and $D$ decrease whereas at the same time both $A$ and $C$ increase. See
main text.
}
\label{fig3}
\end{figure}
%%%%%%%%%%%%%%%%%%%%%%%%%%%%%%%%%%%%%%%%%%%%%%

\begin{figure}
\vglue 5mm
\begin{center}
\includegraphics[width=6.cm,angle=0]{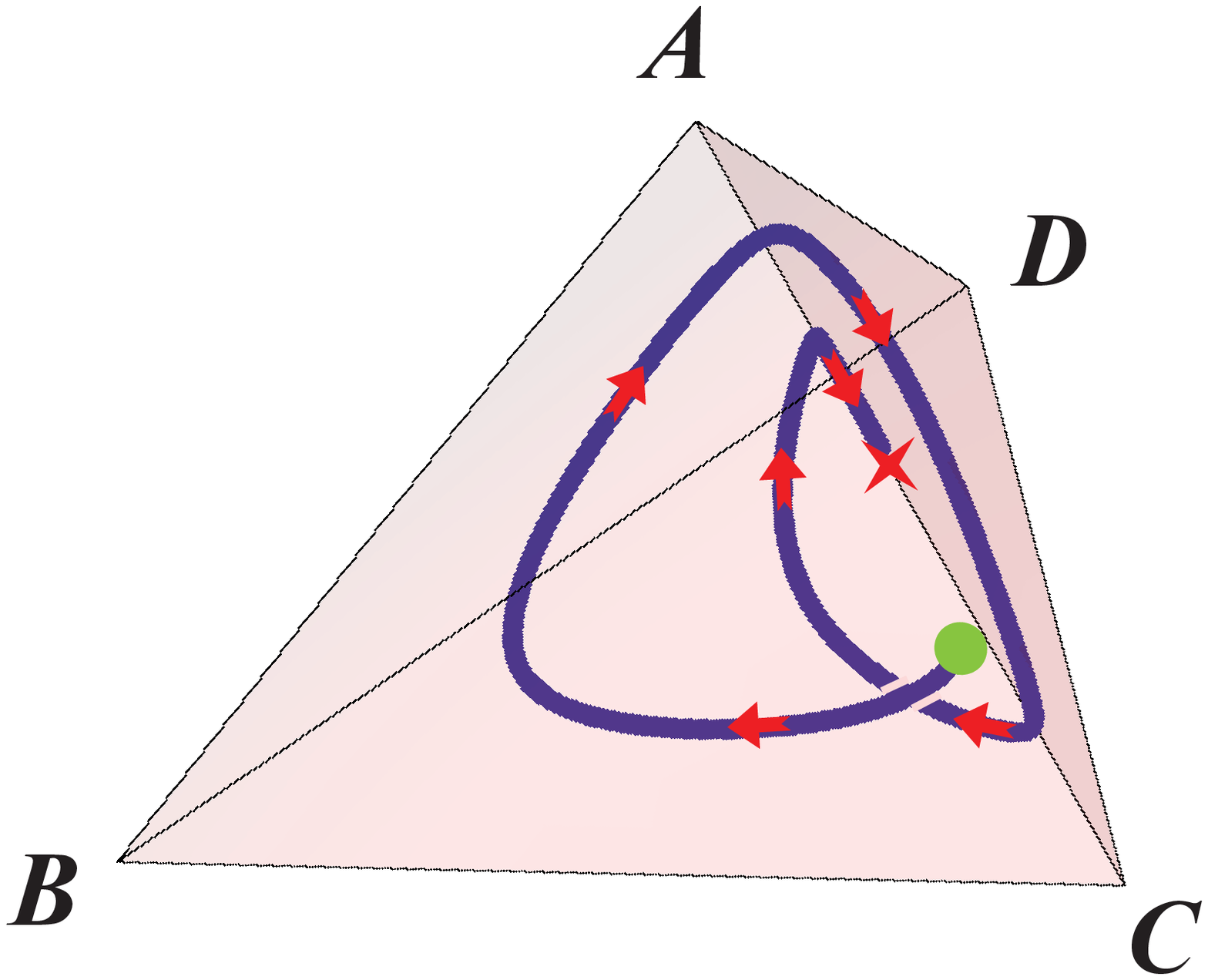}
\end{center}
\caption{(Color online) Typical orbit for $\lambda >0$ that starts
at the solid green circle and spirals towards an absorbing
state, indicated by the red cross on the $a$-$c$ edge.
The data shown here have been obtained for rates
$k_a = 0.45$, $k_b = 0.33$, $k_c =0.14$, $k_d =0.08$ and
initial fractions $A = 0.02$, $B = 0.10$, $C = 0.48$, $D =0.40$.
}
\label{fig4}
\end{figure}

{}From eqns. (\ref{lnAC},\ref{lnBD}), we see that the rise and fall of each
species are associated with the relative values of the opposing pairs. For
example, we can display a line in the $B$-$D$ plane ($k_aB=k_dD$, dashed
line denoted by $\tilde{a}$ in Fig. \ref{fig3}b, on one
side of which $A$ increases (shaded region, yellow online) and on the
other, $A$ decreases. Of course, this `line' is actually a plane in
our configuration space tetrahedron. When the orbit crosses this plane, $A$
reaches a (local) extremum and turns around. A similar line ($k_bB=k_cD$; $%
\tilde{c}$ in Fig. \ref{fig3}b) can be plotted for the `divide' between increasing
(shaded region, green online) and decreasing $C$. The other set of lines, $%
\tilde{b}$ and $\tilde{d}$ in the $A$-$C$ plane, are shown in Fig. \ref{fig3}a. The
significance of $\lambda $ is now clear. For $\lambda =0$, $\tilde{a}=\tilde{%
c}$, so that there are {\em no regions} where both $A$ and $C$ increase or
decrease together. Meanwhile, we also have $\tilde{b}=\tilde{d}$ so that $B$
and $D$ also share such a property. By contrast, if $\lambda >0$, a `gap'
opens between these pairs of lines in such a way that in the region between $%
\tilde{a}$ and $\tilde{c}$, both $A$ and $C$ increase (e.g., overlap of
shaded regions in Fig. \ref{fig3}b, a solid `wedge' in the tetrahedron). On the other
hand, in the region between $\tilde{b}$ and $\tilde{d}$, both $B$ and $D$
decrease (unshaded region in Fig. \ref{fig3}a). As a result, when the system gets
into this domain (intersection of doubly-shaded and unshaded regions within
our tetrahedron -- i.e., an `irregular tetrahedron'), $B$ and $D$ will
monotonically decrease, exponentially at late times. Simultaneously, $A$ and 
$C$ will monotonically increase, towards a final point $\left(
A_f,C_f\right) $ on the $a$-$c$ edge (solid diagonal line in Fig. \ref{fig3}a, red on
line). In particular, the system will end on the heavy solid segment (red on
line) of this line. In Fig. \ref{fig4}, we illustrate these
features with such a ($\lambda >0$) case, showing a typical open orbit
(starting at the solid circle, green online), spiralling towards an absorbing
state on the $a$-$c$ edge (cross, red online).

%%%%%%%%%%% FIG 4  %%%%%%%%%%%%%%%%%%%%%%%%%%%%%%%
%%%%%%%%%%%%%%%%%%%%%%%%%%%%%%%%%%%%%%%%%%%%%%

To appreciate such monotonic behavior better, we present here a unique
trajectory. Instead of being a spiral, it is as `straight as an arrow.' A
system started on this (straight) line will move along it for all $t$. In
fact, as $\lambda \rightarrow 0$, this line becomes the fixed line (\ref{FL}%
) above. Here we simply quote the result. Readers interested in the origins
and general characteristics of such `arrows' will find details in Section
Vb, which is devoted to systems with {\em arbitrary} (even) $S$. We start
with an Ansatz for the fractions 
\begin{equation}
\bar{A}h\left( t\right) ,\,\,\bar{B}\left[ 1-h\left( t\right) \right] ,\,%
\bar{C}h\left( t\right) ,\,\,\bar{D}\left[ 1-h\left( t\right) \right]
\label{arrow}
\end{equation}
where $\bar{A},\bar{B},\bar{C},\bar{D}$ are constants. Clearly, these points
lie on a straight line in the tetrahedron. For $h\in \left[ 0,1\right] $,
this line joins the points $\left( \bar{A},0,\bar{C},0\right) $ and $\left(
0,\bar{B},0,\bar{D}\right) $ on the $a$-$c$ and $b$-$d$ edges, respectively.
Inserting these into eqns. (\ref{ABCD eqns 1},\ref{ABCD eqns 2}), we find
that they can be solved provided 
\[
\partial _t\ln h=\omega \left( 1-h\right) ;\quad \partial _t\ln \left(
1-h\right) =-\omega h 
\]
and 
\[
\omega =k_a\bar{B}-k_d\bar{D}=k_c\bar{D}-k_b\bar{B}=k_a\bar{A}-k_b\bar{C}=k_c%
\bar{C}-k_d\bar{A}\, 
\]
Thus, the result is 
\begin{equation}
\left( \bar{A},\bar{B},\bar{C},\bar{D}\right) =\frac{\left(
k_b+k_c,k_c+k_d,k_a+k_d,k_a+k_b\right) }{k_a+k_b+k_c+k_d}  \label{A-bar}
\end{equation}
so that, explicitly, 
\begin{eqnarray}
\omega &=&\frac{k_a\left( k_b+k_c\right) -k_b\left( k_a+k_d\right) }{%
k_a+k_b+k_c+k_d} \\
&=&\frac \lambda {k_a+k_b+k_c+k_d}\,\,.  \label{omega}
\end{eqnarray}
Since $h$ satisfies $dh/dt=\omega h\left( 1-h\right) $, we may choose $%
h\left( 0\right) =1/2$ and write the evolution along this straight
trajectory explicitly: 
\begin{equation}
h\left( t\right) =\frac{e^{\omega t}}{1+e^{\omega t}}  \label{h(t)}~.
\end{equation}
Note that, indeed, $h\in \left[ 0,1\right] $ for all $t$. Further, as $%
\lambda \rightarrow 0$, so does $\omega $ and $h$ become `frozen,' playing
the role of $\gamma $ in (\ref{FL}). Of course, in this limit, we easily see
that, e.g., the ratio $\bar{A}/\bar{C}=\left( k_b+k_c\right) /\left(
k_a+k_d\right) $ becomes $k_b\left( 1+k_d/k_a\right) /\left( k_a+k_d\right) $
which is $k_b/k_a=A^{*}/C^{*}$. It is reassuring that this `arrow'
trajectory for $\lambda \neq 0$ is continuously linked to the fixed line in $%
\lambda =0$. Apart from this special line, all orbits (which we found
numerically) spiral like a corkscrew, to various degrees, between the $a$-$c$
and $b$-$d$ edges. Classifying the characteristics of these spirals, a task
well suited for future studies, will be highly non-trivial.

\begin{figure}[h]
\vglue 5mm
\begin{center}
\includegraphics[width=8.2cm,angle=0]{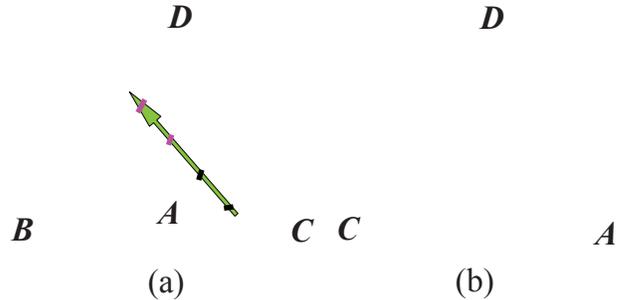}
\end{center}
\caption{(Color online) A typical `arrow' and spiral, with
$\lambda = -0.0273$, from rates 
$( k_a , k_b , k_c , k_d ) = (0.35,0.42,0.09,0.14)$.
(a) Starting at the symmetry point (all fractions being $0.25$), the forward
spiral is shown in magenta [grey]. The $t\rightarrow -\infty $
branch is shown as a black spiral. If started anywhere on the green arrow, the
system will follow the arrow to the $b$-$d$ edge. 
(b) A different perspective of (both branches of) the same spiral.
}
\label{fig5}
\end{figure}

Whether we consider the arrow or spiraling trajectories, we remark that the
time-reversed evolution (starting from any initial point) is also of
interest. In this case, the opposite will occur: For systems with $\lambda >0
$, say, $Q$ will decrease monotonically and the system `ends up' somewhere
on the opposite ($b$-$d$) edge. In this sense, the full dynamics for $%
\lambda \neq 0$ can be regarded as a family of spirals around an arrow,
bridging the two opposite edges 
%(illustrated in Fig. \ref{fig5}a, with a single
%spiral). 
The initial condition simply serves to
pick out the appropriate orbit, while the evolution consists of following it
to an absorbing edge. 
%Shown in Fig. \ref{fig5}b is the same spiral as in Fig. \ref{fig5}a, but
%with {\em two} branches (solid vs. dashed lines, red vs. blue online)
Shown in Fig. \ref{fig5} is such a spiral, but
with {\em two} branches (black line vs. magenta [grey] line)
corresponding to $t\rightarrow \pm \infty $ (starting at the same point).
Underlying this dynamics appears to be a kind of `duality' symmetry
(associated with $\lambda ,t\Leftrightarrow -\lambda ,-t$). Exploring this
aspect is likely a worthy pursuit, but clearly beyond the scope of this
paper.

%%%%%%%%%%%%%%%%%%%%%%%%%%%%%%%%%%%%%%Fig. 5%%%%%%%%%%%%%%%%%%%%%%%%%%%%%
%%%%%%%%%%%%%%%%%%%%%%%%%%%%%%%%%%%%%%%Fig. 5%%%%%%%%%%%%%%%%%%%%%%%%%%%%%^M

\section{Implications for stochastic evolution}

When compared to the evolution of a fully stochastic 4SS with finite $N$,
the predictions of the deterministic MFT can be a good approximation for the
average behavior for cases with large $N_m$'s. Of course, MFT has serious
limitations as well, apart from its inherent inability to describe
fluctuations. The most glaring discrepancy lies with issues concerning
extinction. Examples which readily come to mind are: What is the eventual
extinction probability of a species? At what time is a species most likely
to go extinct? And what is the average time for extinction? Since the
fractions in MFT never vanish in finite time, none of these can be answered.
Nevertheless, MFT can provide valuable insights \cite{Ber09}, some of which
will be provided here. Extensive studies of the stochastic system and
comparisons with MFT are beyond the scope of this article and will be
presented in a subsequent publication \cite{CDPZ-JSTAT}.

Let us consider first the $\lambda =0$ case, where only closed orbits
prevail and all species survive in the MFT. This behavior resembles that in
a 3SS. As pointed out in \cite{Ber09}, the closest approaches a closed loop
makes with each edge (of the configuration space triangle) are good
indicators of the survival probabilities. We should caution the reader that,
in the 3SS triangle, an edge is associated with the \emph{survival} of a
single species, as well as the \emph{extinction} of one other species. Thus,
it is natural to associate a minimum distance to an edge ($\lambda _m$, in
the notation of \cite{Ber09}) with the survival of species $m$ (instead of
the extinction of species $m-1$). In our case, the extinction of a species
induces the survival of the opposing partner-pair. Thus, we find it more
convenient to use the language of extinction of a species rather than likely
survival of a pair. In other words, the closest approach to one of the four
faces (in our tetrahedron) is a good indicator of the extinction probability
of the associated species. For example, $A_{-}$ is the closest a loop comes
within the face $A=0$, which corresponds to $a$ being extinct. Therefore, we
can expect a species associated with the smallest in the set $\left(
A_{-},B_{-},C_{-},D_{-}\right) $ to vanish first. In contrast to the 3SS
case, the rates alone do not provide a useful criterion for extinction in
our case, as we will show next.

In the 3SS, the initial conditions identifies a closed orbit and is
associated with the invariant $R=A^{k_b}B^{k_c}C^{k_a}$. It is easy to find
the closest approaches $\left( A_{-},B_{-},C_{-}\right) $ and to compute
their ratios. For orbits with $R\ll 1$, these ratios are well approximated
by 
\begin{eqnarray}
A_{-}/B_{-} &\propto &R^{1/k_b-1/k_c}\,\,~,  \label{3ssRatios} \\
B_{-}/C_{-} &\propto &R^{1/k_c-1/k_a}~, \\
C_{-}/A_{-} &\propto &R^{1/k_a-1/k_b}~.
\end{eqnarray}
Notably, the proportionality coefficients depend only on the rates and \emph{%
not} on $R$. Therefore, given a set of rates and a large population ($N\gg 1$%
), it is possible to arrive, through stochastic effects, at orbits where one
of the fractions $N_m/N\rightarrow 1/N\ll 1$. Considering orbits with
smaller and smaller $R$, we see that the above ratios will vanish or
diverge, depending on the rates alone. For example, if $k_a<k_b,k_c$ , then $%
C_{-}$ will be the smallest as $R\rightarrow 0$, so that $c$ is be the most
likely to go extinct first. This argument leads to the conclusion in \cite
{Ber09}: Extinction/survival probabilities are either \emph{zero} or \emph{%
unity}, in the limit of large $N$. By contrast, two invariants ($%
f=A^{k_b}C^{k_a},g=B^{k_d}D^{k_a}$) are associated with the closed loops in the 4SS. As
a result, similar ratios of the closest approaches are not simple functions
as in (\ref{3ssRatios}). Let us examine this issue further, by studying an
example. We will again consider cases with small closest approaches, so that
we can ignore the first term in eqn. (\ref{A-hat eqn}) for finding, say, 
$A_{-}$. The results are 
\begin{eqnarray*}
A_{-} &\simeq &f^{1/k_b}\left[ 1-K_Ag^{1/\left( k_a+k_d\right) }\right]
^{-k_a/k_b} \\
C_{-} &\simeq &f^{1/k_a}\left[ 1-K_Ag^{1/\left( k_a+k_d\right) }\right]
^{-k_b/k_a} \\
B_{-} &\simeq &g^{1/k_d}\left[ 1-K_Df\,\,^{1/\left( k_a+k_b\right) }\right]
^{-k_a/k_d} \\
D_{-} &\simeq &g^{1/k_a}\left[ 1-K_Df\,\,^{1/\left( k_a+k_b\right) }\right]
^{-k_d/k_a}
\end{eqnarray*}
where $K_{A,D}$ are given by (\ref{KA},\ref{KD}) and are \emph{independent}
of $f,g$. Even the ratios of each partner pair are not so clear-cut as in (%
\ref{3ssRatios}), since the `other invariant' enters in a nontrivial way.
For example, $g$ appears in the ratio 
\[
\frac{A_{-}}{C_{-}}\simeq f^{1/k_b-1/k_a}\left[ 1-K_Ag^{1/\left(
k_a+k_d\right) }\right] ^{k_b/k_a-k_a/k_b}\,\,.
\]
Needless to say, `cross-ratios' such as $A_{-}/B_{-}$ are even more
intractable. In a stochastic evolution, $f$ and $g$ presumably wander
separately, so that it is \emph{a priori} unclear which pair will win. Since
the qualitative picture is already quite complex, we will not pursue the
route in \cite{Ber09} where more quantitative arguments concerning scaling
can be fruitful. Instead, devoting efforts towards the study of the full
stochastics, along the lines of \cite{Dob10}, is likely to be more worthwhile.

Turning to systems with $\lambda \neq 0$, we can reach similar qualitative
conclusions concerning how well the MFT predicts the behavior of the
stochastic model. Here, $Q$ grows/decays exponentially and which pair
survives is determined only by the sign of $\lambda $. Phrased in stark
terms, this conclusion states that neither its magnitude nor the initial
configuration (as long as they lie within the tetrahedron) plays a role.
Clearly, this cannot be true for stochastically evolving, finite systems.
{}From our simulations \cite{CDPZ-JSTAT}, we glean another maxim: ``The prey of
the prey of the strongest is \emph{quite likely} to survive.'' Although this
principle appears to be the same as the one posed above (albeit a
double-negative version), the consequences for the end state can be
drastically different. In particular, both $\left| \lambda \right| $ and the
initial conditions, as well as the total population size, can reverse the
predictions of MFT. Let us illustrate with the example in \cite{CDPZ-epl} with
`extreme' rates: $k_m=\left( 0.1,0.001,0.1,0.7999\right) $. Since $\lambda
>0 $, $a$-$c$ is predicted to win. When the initial population is set at $%
N_m=\left( 100,700,100,100\right) $, 90\% of the runs indeed end with the $a$%
-$c$ pair. However, for a symmetric starting point of $N_m=\left(
100,100,100,100\right) $, 97\% of the runs end with $b$ as the sole
survivor, \emph{despite }the initial $N_b$ being much smaller! 
To accommodate these two very different outcomes, we apply the two different
maxims. As $b$ is the weakest and $d$ is the strongest, each is the prey of
the other's prey. In one case, the first maxim lead us to $D$ vanishing
first, leaving us with $a$-$c$ coexistence. In the second scenario, $d$
consumes $a$ so rapidly that $A$ vanishes first and $b$ can only increase.
But $b$ is slow to consume $c$, so that $d$ goes extinct next. As a result, $%
b$ is the only survivor (rather than one of the more numerous states of $b$-$%
d$ coexistence). 
To turn these qualitative
considerations into quantitative conclusions, concerning the detail
interplay between the rates and the initial configuration, will be a serious
challenge.

To summarize this section, let us reiterate that, while MFT cannot predict 
\emph{all} eventualities of a finite stochastic system, it can provide some
hints. Specifically, the proximity of an MFT trajectory to an absorbing face
should be a good guide for the extinction probabilities of the associated
species. For the 3SS, this guide led to very successful predictions, i.e.,
survival of the weakest. This success can be traced to the presence of only
a few degrees of freedom, so that, along with the existence of the invariant 
$R$, the rates alone determine the relative proximity of the orbits to the
three extinction lines. By contrast, the 4SS is sufficiently more complex
that similar conclusions cannot be drawn.

\section{Systems with $S$ species}

In this section, we present extensions of our findings to a system of $N$
individuals consisting of an \emph{arbitrary} number, $S$, of cyclically
competing species.
This generalization provides some insight into the
general structure of the dynamics in such systems. In particular, these
considerations lead us to the deeper origins of the existence of collective
variables like $R$ and $Q$, as well as fixed points and `arrows.' For any
odd $S$, we show that there is precisely one (non-trivial) fixed point and
one $R$-like invariant. By contrast, for even $S$, a $Q$-like quantity
(evolving simply as an exponential) and an `arrow' trajectory always exist.
Regarding the system as two opposing teams, we show that the winners always
have the larger rates-product (e.g., $a$-$c$ if $k_ak_c$ is larger). When
the competition is neutral (equal rate-products), there will always be a
fixed line and \emph{two} $R$-like invariants.

Following Section II, we now consider the species $x_m$ ($m=1,\cdots ,S$)
interacting cyclically: 
\begin{equation}
x_m+x_{m+1}\stackrel{p_m}{\rightarrow }2x_m
\end{equation}
(with $x_{S+1}=x_1$). The rate equations for the \emph{fractions} $X_m\equiv
N_m/N$ take the form 
\begin{eqnarray}
\partial _tX_m=\left[ k_mX_{m+1}-k_{m-1}X_{m-1}\right] X_m~.  \label{X eqn}
\end{eqnarray}
Of course, $\sum_mX_m=1$, so that our configuration space is an $S-1$
simplex (a hyper-tetrahedron). As in Section III, a better approach is to
cast eqn. (\ref{X eqn}) as 
\begin{eqnarray}
\partial _t\left( \ln X_m\right) =k_mX_{m+1}-k_{m-1}X_{m-1} ~. \label{lnX eqn}
\end{eqnarray}
Despite the non-linearity, we will find it convenient to exploit the
notation of vectors and matrices. Thus, we rewrite this equation as 
\begin{equation}
\partial _t\overrightarrow{\ln X}=\Bbb{K}\vec{X}  \label{U eqn}
\end{equation}
where 
\begin{equation}
\Bbb{K}=\left( 
\begin{array}{cccccc}
0 & k_1 & 0 & ... & 0 & -k_S \\ 
-k_1 & 0 & k_2 & ... & 0 & 0 \\ 
0 & -k_2 & 0 & ... & 0 & 0 \\ 
\vdots & \vdots & \vdots & ... & \vdots & \vdots \\ 
0 & 0 & 0 & ... & 0 & k_{S-1} \\ 
k_S & 0 & 0 & ... & -k_{S-1} & 0
\end{array}
\right)  \label{K def}
\end{equation}
is an antisymmetric, cyclic matrix and $\vec{X}$ is a vector with elements $%
X_m$. Introducing 
\[
\vec{E}=\left( 
\begin{array}{ccccc}
1 & 1 & ... & 1 & 1
\end{array}
\right) 
\]
we see that $\sum_mX_m=1$ is simply $\vec{E}\cdot \vec{X}=1$. The
significance of eqn. (\ref{U eqn}) is clear: It highlights the difference
between having a singular $\Bbb{K}$ or not. Such properties are best
explored by considering systems with odd/even $S$ separately. Indeed, the
major differences between the 3SS and 4SS can be understood in the light of
this even/odd distinction.
(Odd-even effects were briefly discussed in \cite{Sat02}
for systems on a two-dimensional lattice).

\subsection{Systems with odd $S$}

Antisymmetric matrices have a simple property: If $\kappa $ is an
eigenvalue, then so is $-\kappa $. Since $S$ is odd, there must be at least
one eigenvalue which is zero (the rest being pairs of opposite signs).
Indeed, our $\Bbb{K}$ has only one such eigenvalue, as shown in Appendix B,
so that its null space is one-dimensional. The associated \emph{right}
eigenvector, $\vec{\zeta}$, satisfies $\Bbb{K}\vec{\zeta}=0$. All its
elements, $\zeta _m$, can be chosen to be positive, since they obey $%
k_{m-1}\zeta _{m-1}=k_m\zeta _{m+1}$. Of course, when normalized
appropriately, it becomes the fixed point population: 
\begin{equation}
\vec{X}^{*}=\frac{\vec{\zeta}}{\vec{E}\cdot \vec{\zeta}} ~. \label{X*}
\end{equation}
Since $\Bbb{K}$ is antisymmetric, $\vec{\zeta}$ is also the \emph{left}
eigenvector ($\vec{\zeta}\cdot \Bbb{K}=0$). Thus, we arrive at $\partial _t%
\vec{\zeta}\cdot \overrightarrow{\ln X}=0$, which leads us to define an
invariant 
\begin{equation}
\mathcal{R}\equiv \prod_mX_m^{\zeta _m}\,\,.  \label{curly R}
\end{equation}
Note that $\vec{\zeta}$ is determined up to an overall constant, which
simply corresponds to $\mathcal{R}$ being fixed (up to an overall power).
This is the generalization of $R$ in the 3SS. Together with $\vec{E}\cdot 
\vec{X}=1$, this invariant defines a compact, $S-2$ dimensional manifold on
which the orbit must lie. We find it remarkable that $\vec{\zeta}$ plays a
`dual' role, serving to pinpoint both the fixed point $\vec{X}^{*}$ and to
define the invariant $\mathcal{R}$. An interesting question naturally
arises: Does a deeper connection between these apparently unrelated
quantities exist?

\subsection{Systems with even $S$}

We may regard such systems as the competition between two `teams,' each with 
$s\equiv S/2$ species. In particular, this aspect becomes transparent if we
reorder our species to 
\[
X_1,X_3,...,X_{S-1},X_2,...,X_S 
\]
and define `team variables' 
\[
Y_\ell \equiv X_{2\ell -1};\quad Z_\ell \equiv X_{2\ell } 
\]
with $\ell =1,...,s$. In terms of these, the matrix $\Bbb{K}$ takes the form 
\begin{equation}
\Bbb{K}=\left( 
\begin{array}{cc}
0 & \Bbb{M} \\ 
-\Bbb{M}^{\text{T}} & 0
\end{array}
\right)  \label{K-M}
\end{equation}
and eqn. (\ref{U eqn}) becomes 
\begin{equation}
\partial _t\left( 
\begin{array}{c}
\overrightarrow{\ln Y} \\ 
\overrightarrow{\ln Z}
\end{array}
\right) =\left( 
\begin{array}{cc}
0 & \Bbb{M} \\ 
-\Bbb{M}^{\text{T}} & 0
\end{array}
\right) \left( 
\begin{array}{c}
\vec{Y} \\ 
\vec{Z}
\end{array}
\right)  \, .\label{YZ eqn}
\end{equation}
Of course, the full space and `team space' have different dimensions ($S$
vs. $s$), but using the same notation ($\Bbb{K}$, $\Bbb{M}$, $\vec{X}$, $%
\vec{Y}$, etc.) should not cause much confusion. Note that eqn. (\ref{YZ eqn}%
) exposes a simplectic structure that is typical in the `competition of two
(sets of) variables.'

Careful accounting of the $s\times s$ matrix $\Bbb{M}$ leads to: 
\begin{equation}
\Bbb{M}=\left( 
\begin{array}{ccccc}
k_1 & 0 & ... & 0 & -k_S \\ 
-k_2 & k_3 & ... & 0 & 0 \\ 
0 & -k_4 & ... & 0 & 0 \\ 
\vdots & \vdots & ... & \vdots & \vdots \\ 
0 & 0 & ... & -k_{S-2} & k_{S-1}
\end{array}
\right)  \label{M def}
\end{equation}
Note that its structure is quite different from $\Bbb{K}$, with the (even)
odd rates lining the (sub-)diagonal. Whether $\Bbb{K}$ is singular or not is
just controlled by $\Lambda \equiv \det \Bbb{M}$. A straightforward
computation leads to 
\[
\Lambda =\prod_{odd\,\,m}k_m-\prod_{even\,\,m}k_m 
\]
regardless of whether $s$ is even or odd. Clearly, $\Lambda $ is the
generalization of $\lambda $ in the 4SS and its sign determines which team
wins. If $\Lambda \neq 0$, $\Bbb{K}^{-1}$ exists so that $\partial _t\left[ 
\vec{E}\cdot \Bbb{K}^{-1}\overrightarrow{\ln X}\right] =\vec{E}\cdot \vec{X}%
=1$. Denoting $\sum_j\left( K^{-1}\right) _{jm}$ by $\sigma _m$ and defining 
\begin{equation}
\mathcal{Q}\equiv \exp \left[ \vec{E}\cdot \Bbb{K}^{-1}\overrightarrow{\ln X}%
\right] =\prod_mX_m^{\sigma _m}  \label{curly Q}
\end{equation}
we see that it evolves trivially: $\mathcal{Q}\left( 0\right) e^t$. Of
course, it is impossible to form such teams in systems with odd $S$, there
is no $\mathcal{Q}$-like variable, and this simple scenario is invalid.

Though obviously related to $Q$ in the 4SS, $\mathcal{Q}$ is less
transparent, since $e^t$ obscures the $\Lambda $ dependence of our system!
To find the generalization of $Q$ and to shed light on the $\Lambda
\rightarrow 0$ limit, we exploit the team variables $\left( \vec{Y},\vec{Z}%
\right) $ and consider $\Bbb{W}$, the cofactors of $\Bbb{M}$. Unlike $\Bbb{M}%
^{-1}$, 
\begin{equation}
\Bbb{W}=\Lambda \Bbb{M}^{-1}  \label{WLM}
\end{equation}
is well behaved even when $\Lambda =0$. Multiplying eqn. (\ref{YZ eqn}) by 
\begin{equation}
\left( 
\begin{array}{cc}
0 & -\Bbb{W}^{\text{T}} \\ 
\Bbb{W} & 0
\end{array}
\right) 
\end{equation}
we have 
\begin{equation}
\partial _t\left( 
\begin{array}{c}
-\Bbb{W}^{\text{T}}\,\overrightarrow{\ln Z} \\ 
\Bbb{W}\,\overrightarrow{\ln Y}
\end{array}
\right) =\Lambda \left( 
\begin{array}{c}
\vec{Y} \\ 
\vec{Z}
\end{array}
\right) ~.  \label{WU}
\end{equation}
Now, a tedious computation shows that all the matrix elements of $\Bbb{W}$
are positive (sums of products of $s-1$ rates, e.g., $k_b+k_c$ in the $s=2$
case above). Let us define the following \emph{positive} quantities 
\begin{equation}
q_\ell \equiv \sum_nW_{n\ell };\quad r_n\equiv \sum_\ell W_{n\ell }
\label{qr}
\end{equation}
which are readily recognized as $\vec{E}\cdot \Bbb{W}$ and $\vec{E}\cdot 
\Bbb{W}^{\text{T}}$, respectively. Thus, by taking the scalar product of $%
\vec{E}$ with eqn. (\ref{WU}), we are led to the combination 
\begin{equation}
Q\equiv \exp \left[ \vec{q}\cdot \overrightarrow{\ln Y}-\vec{r}\cdot 
\overrightarrow{\ln Z}\right] =\left. \prod_\ell Y_\ell ^{q_\ell }\right/
\prod_\ell Z_\ell ^{r_\ell }  \label{Q def}
\end{equation}
which evolves according to 
\begin{equation}
Q\left( t\right) =Q\left( 0\right) e^{\Lambda t}\,\,.  \label{Q evolve}
\end{equation}
To help the readers, let us connect these results to the 4SS case
explicitly. There, 
\[
\left( 
\begin{array}{c}
\vec{Y} \\ 
\vec{Z}
\end{array}
\right) =\left( 
\begin{array}{c}
A \\ 
C \\ 
B \\ 
D
\end{array}
\right) 
\]
and 
\[
\Bbb{M}=\left( 
\begin{array}{cc}
k_a & -k_d \\ 
-k_b & k_c
\end{array}
\right) ;\quad \Lambda =\lambda ;\quad \Bbb{W}=\left( 
\begin{array}{cc}
k_c & k_d \\ 
k_b & k_a
\end{array}
\right) 
\]
so that all the conclusions reached in Section III are recovered.

As in the 4SS, the orbits are, in general, analytically inaccessible.
However, the `straight arrow' trajectory can be determined explicitly, while
the time dependence along it remains the same for any number of species. In
particular, we begin with the most trivial case: $s=1=S/2$: 
\begin{equation}
\partial _tX_1=k_1X_1X_2;\quad \partial _tX_2=-k_1X_1X_2  \label{X1X2}
\end{equation}
so that $\Lambda $ is just $k_1$. and $Q=X_1/X_2\propto e^{k_1t}$. Since $%
X_2=1-X_1$, the solution is trivial, being the same form as (\ref{h(t)})
with $k_1$ in place of $\omega $. As $t$ runs to $-\infty $ or $+\infty $,
it reaches the two terminals $X_1=1$ or $X_2=1$. Turning to general $S$, we
seek such a straight line, bridging the terminal points ($\bar{Y}_\ell $ and 
$\bar{Z}_\ell $) in the subspaces of each team, that can serve as a $t$%
-dependent trajectory. Making the same Ansatz as in eqn. (\ref{arrow}), let
us consider 
\[
\bar{Y}_\ell h\left( t\right) ,\,\bar{Z}_\ell \left[ 1-h\left( t\right)
\right] 
\]
which runs from $\left( \bar{Y}_\ell ,0\right) $ to $\left( 0,\bar{Z}_\ell
\right) $. We will also assume the same $h\left( t\right) $ as in (\ref{h(t)}%
), except for $\omega $ being replaced by $\Omega $, a parameter to be
determined. Inserting these into the left of eqn. (\ref{YZ eqn}), we find 
\begin{equation}
\left( 
\begin{array}{c}
\vec{E}\,\partial _t\ln h \\ 
\vec{E}\,\partial _t\ln \left[ 1-h\right] 
\end{array}
\right) =\Omega \left( 
\begin{array}{c}
\vec{E}\left( 1-h\right)  \\ 
-\vec{E}\,h
\end{array}
\right)  ~. \label{arrow1}
\end{equation}
Meanwhile the right of eqn. (\ref{YZ eqn}) is 
\begin{equation}
\left( 
\begin{array}{c}
\Bbb{M}\,\vec{Z} \\ 
-\Bbb{M}^{\text{T}}\,\vec{Y}
\end{array}
\right) =\left( 
\begin{array}{c}
\vec{\Xi}\left( 1-h\right)  \\ 
-\vec{\Upsilon}\,h
\end{array}
\right)   \label{arrow2}
\end{equation}
where the components of $\vec{\Upsilon}$ and $\vec{\Xi}$ on the right are,
respectively, 
\begin{equation}
\sum_\ell M_{\ell n}\bar{Y}_\ell \quad \text{and}\quad \sum_\ell M_{n\ell }%
\bar{Z}_\ell ~.  \label{MZ+MY}
\end{equation}
Setting expressions (\ref{arrow1},\ref{arrow2}) equal, we find 
\begin{eqnarray*}
\bar{Y}_n &=&\Omega \sum_\ell \left( M^{-1}\right) _{\ell n} ~,\\
\bar{Z}_n &=&\Omega \sum_\ell \left( M^{-1}\right) _{n\ell } ~.
\end{eqnarray*}
Exploiting (\ref{WLM},\ref{qr}), we have 
\begin{eqnarray}
\bar{Y}_n &=&\frac \Omega \Lambda \sum_\ell W_{\ell n}=\frac \Omega \Lambda
q_n ~, \label{Ybar} \\
\bar{Z}_n &=&\frac \Omega \Lambda \sum_\ell W_{n\ell }=\frac \Omega \Lambda
r_n  ~. \label{Zbar}
\end{eqnarray}
Now, all details for our `arrow' condition are fixed, by imposing $\sum_n%
\bar{Y}_n=1$ etc. The result is 
\begin{equation}
\Omega =\Lambda \left/ \sum_{n,\ell }W_{\ell n}\right.   \label{OMEGA}
\end{equation}
i.e., the generalization of (\ref{omega}).

We end with some remarks on the cases with $\Lambda =0$. Although properties
of these can be found by taking the $\Lambda \rightarrow 0$ limit of the
above, let us provide a more direct route. Since $S$ is even, a zero
eigenvalue of $\Bbb{K}$ is doubly degenerate and the null space is
two-dimensional (at the least). Given the form (\ref{K-M}), we see that this
null space is determined by the \emph{right} eigenvectors of $\Bbb{M}$ and $%
\Bbb{M}^{\text{T}}$: 
\[
\Bbb{M}\,\vec{\xi}=0=\Bbb{M}^{\text{T}}\vec{\eta}\,~.
\]
We can impose the constraint and define fixed \emph{points} within the space
of each `team': 
\[
Y_\ell ^{*}=\xi _\ell \left/ \sum_{n=1}^s\xi _n\right. ;\quad Z_\ell
^{*}=\eta _\ell \left/ \sum_{n=1}^s\eta _n\right. ~.
\]
Joining these is a fixed line in the full $S-1$ simplex, that can be
parametrized in the same manner as in eqn. (\ref{FL}) of the 4SS: 
\[
\left( 
\begin{array}{c}
\vec{Y}^{*} \\ 
\vec{0}
\end{array}
\right) \gamma +\left( 
\begin{array}{c}
\vec{0} \\ 
\vec{Z}^{*}
\end{array}
\right) \left( 1-\gamma \,\,\right) \,\,.
\]
Meanwhile, the \emph{left }eigenvectors of $\Bbb{K}$, are expected to play a
`dual' role, as in the odd $S$ cases. Of course, these can be formed from $%
\vec{\eta}$ and $\vec{\xi}$, which are the left eigenvector of $\Bbb{M}$ and 
$\Bbb{M}^{\text{T}}$, respectively. The results are $\left( \vec{\eta},\vec{0%
}\right) $ and $\left( \vec{0},\vec{\xi}\right) $, which can now be
exploited to define the generalizations of $f$ and $g$, see eq. (\ref{f,g}): $%
\vec{\eta}\cdot \overrightarrow{\ln Y}$ and $\vec{\xi}\cdot \overrightarrow{%
\ln Z}$. Not surprisingly, these are intimately related to the numerator and
the denominator in expression (\ref{Q def}). Note that the sum 
$\vec{\eta}\cdot \overrightarrow{\ln Y}+\vec{\xi}\cdot \overrightarrow{\ln Z}$ 
is also an invariant. 
Exponentiating it will produce a quantity which has the
appearance of $\mathcal{R}$, i.e., products of only positive powers of $X_m$%
. For systems with all interaction rates being equal, such an invariant
takes an appealing form: $X_1X_2...X_S$ (for any $S$). Finally, we note that 
$\Bbb{M}$ has no other zero eigenvalues (shown in Appendix B), so that there
are no other invariants. Thus, the orbits lie in a compact, $S-3$
dimensional manifold in general. Of course, for $S=4$, the 1-dimensional
manifolds are simply closed loops, for which many properties are explicitly
obtained in Section III. For $S>4$, we have not found similar explicit
characteristics.

\section{Summary and outlook}
In an effort to understand better the counter-intuitive phenomenon of
``survival of the weakest,'' which was discovered in recent studies of a
well-mixed system involving three cyclically competing species, we
investigate a system with four species. In a previous letter \cite{CDPZ-epl},
we reported preliminary findings, showing that this behavior does not
persist. Instead, all systems seem to follow an intuitively understandable
principle: ``The prey of the prey of the weakest is most likely to go
extinct first.'' In a 3SS, this species is also the predator of the weakest,
and so, its early demise leads to the weakest being the sole survivor. By
contrast, the players in a 4SS form partner-pairs, much like in the game of
Bridge. As a result, our new maxim hints that the weaker \emph{pair} is not
likely to survive. All these studies consist of two aspects, each of
interest on their own. The first is mean field theory (MFT), i.e.,
non-linear dynamics of a set of deterministic rate equations. The second is
the stochastic evolution of a finite population of $N$ individuals. For the
4SS, we show how the former, our set of rate equations: (\ref{ABCD eqns 1})-%
(\ref{ABCD eqns 4}), arises from the master equation which defines the latter.
This article is devoted to an in-depth exploration of the MFT, i.e.,
solutions to the rate equations.

Since the evolution of the 4SS takes place within a tetrahedron, it is more
complex than the 3SS case. Nevertheless, we are able to obtain considerable
details analytically, mainly through the discovery of hidden symmetries. Not
surprisingly, far fewer exact results of the \emph{stochastic} aspect are
attainable. Instead, most insight into this behavior is gained through
computer simulations. Indeed, in some `extreme' cases, we found that the
outcomes contradict the MF prediction completely \cite{CDPZ-epl}. Instead,
these can be understood by a complementary version of the above maxim, namely,
``The prey of the prey of the strongest is most likely to survive.'' A
similar, in-depth investigation of the second aspect (stochastic evolution)
is beyond the scope here, and those results will be reported elsewhere \cite
{CDPZ-JSTAT}.

Unlike the 3SS, in which all MFT orbits are closed loops (characterized by 
$R $) in a plane, the 4SS supports a richer variety of trajectories. The
details depend on the initial conditions and the rates, especially through 
$\lambda =k_ak_c-k_bk_d$. We show that, provided they start within our
tetrahedron, all orbits spiral, to various degrees, into the $a$-$c$ edge 
($\lambda >0$) or the $b$-$d$ edge ($\lambda <0$). There is one exception,
namely, a special straight line bridging the two edges. On it the system
moves monotonically, leading us to refer to it as the `arrow.' In general,
the spirals wrap around this arrow. To describe the motion along the spirals
and the arrow quantitatively, we find a collective variable, $Q$,
which evolves simply as an exponential, see eqns. (\ref{Q-def},\ref{Q-evol}). 
Indeed, letting $t$ run from $-\infty $ to $\infty $, 
we see that these trajectories all start on one edge ($a$-$c$ or $b$-$d$) 
and end on the opposite one. As 
$\left| \lambda \right| \rightarrow0 $, 
the spirals become tighter (lower pitched) so that, for $\lambda =0$,
all orbits are closed loops which resemble the edge of a saddle. Meanwhile,
the `arrow' becomes a line of fixed points, around which all closed orbits
encircle. In addition to $Q$, another invariant emerges. Together, these two,
see eq. (\ref{f,g}), completely specify the saddle shaped loops. Though the
explicit forms are not available, many properties of these loops, e.g.,
maximal extent in principle directions, have been found. Finally, by
extending these considerations to cyclic competition of an \emph{arbitrary}
number, $S$, of species, we can trace the origins of these special
quantities (invariants, collective variables, non-trivial fixed points and
lines, etc.) to general properties of an $S\times S$ antisymmetric, cyclic
matrix, $\Bbb{K}$. The elements of $\Bbb{K}$ are just the rates, so that all
details of these special quantities are explicitly known. Indeed, a recent
study finds that such `hidden symmetries' are present in an even wider class
of systems involving competing species \cite{Zia10}. Of course, as our scope
broadens, fewer explicit results are available.

Let us reiterate that the main forte of MFT is to provide reliable estimates
of the evolution when the population of every species is large. Relying on
continuous fractions (and continuous time), it fares poorly when one or more
species are near extinction and their numbers drop to $O\left( 1\right) $.
Though it fails to predict all eventualities of a finite stochastic system,
it does give us some useful hints. Specifically, the proximity of an MFT
trajectory to an absorbing face should be a good guide for the extinction
probabilities of the associated species. We have not systematically explore
this issue and such an undertaking seems worthwhile.

Beyond such immediate questions, we believe there are interesting avenues
for further research. We end with pointing out a few here. The sheets
labeled by $\tilde{a}$, $\tilde{b}$, etc. are clearly significant, as they
correspond to the turning points of species $a$, $b$, etc. It seems likely
that these are good candidates for Poincar\'{e} sections with which to analyze
our dynamical system. How often orbits pierce these sheets may also serve to
classify the spirals by the total number of turns, $\chi $ (`windings'),
between the end points. Clearly, $\chi $ is zero for the arrow, while our
expectation is that this value will be infinitesimal for a spiral in its
neighborhood. In other words, a perturbative approach is likely to be
successful. Of course, to be quantitative, we should look for a good
coordinate system for configuration space, guided by slow and fast dynamics,
along the lines of action-angle variables in classical mechanics. For the
neutral cases ($\lambda =0$), these have been identified: $f,g,\theta $ in
eqns. (\ref{f,g}, \ref{eq:theta}). 
In the general case, we already have one key variable, $Q$.
It seems quite feasible to find two others, perhaps in analog to a radius
(distance from the arrow) and an angle, $\phi $. Once their $t$-dependence
is established, then the classification of orbits we conjectured would be
realized by $\chi \equiv \phi \left( t=\infty \right) -\phi \left( t=-\infty
\right) $. It is even conceivable that a deeper connection with similar
concepts in scattering theory exists.

\begin{acknowledgments}
We are grateful to E. Frey, K. Mallick, 
S. Redner, B. Schmittmann, E. Sharpe, and Z. Toroczkai for 
illuminating discussions. This work is supported in part by the 
US National Science Foundation through Grants DMR-0705152, 
DMR-0904999 and DMR-1005417.
\end{acknowledgments}

\appendix

\section{Properties of closed orbits}

Quantitative details of a closed orbit can be found through its projection
onto, say, the $\rho _a$-$\rho _d$ plane. Using $A=A_0\rho _a^{1/k_b}$,
etc., the (generically transcendental) equation for this projection is given
by eqn. (\ref{saddle1}): 
\[
A_0\rho _a^{1/k_b}+B_0\rho _d^{-1/k_d}+C_0\rho _a^{-1/k_a}+D_0\rho
_d^{1/k_a}=1\,\,. 
\]
Clearly, this is just one of many equivalent representations. We begin with
finding the extremal points of this closed loop. Exploiting such points, we
can transform this cumbersome equation into a familiar one, (\ref
{unit
circle}) below. Finally, we will show how the period of these orbits
can be found, explicitly in certain cases.

To find the extremal points, let us first consider the combination involving 
$\rho_a$: 
\[
F\left( \rho_a\right) \equiv A_0\rho _a^{1/k_b}+C_0\rho
_a^{-1/k_a}\,\,. 
\]
As $F$ is convex, it has a unique minimum value, $\stackrel{\vee }{F}$,
which occurs at 
\[
\stackrel{\vee }{\rho }_a\equiv \left[ \frac{k_bC_0}{k_aA_0}\right] ^{\frac{%
k_ak_b}{k_a+k_b}}\,\,, 
\]
and corresponds to 
\[
\stackrel{\vee }{A}=A_0\stackrel{\vee }{\rho }_a^{1/k_b}=\left[
k_b^{k_a}k_a^{-k_a}A_0^{k_b}C_0^{k_a}\right] ^{1/\left( k_a+k_b\right) } ~.
\]
In other words, $F\left( \stackrel{\vee }{\rho }_a\right) =\stackrel{\vee }{F%
}$. Similarly, the other combination, 
\[
G\left( \rho_d\right) \equiv B_0\rho _d^{-1/k_d}+D_0\rho _d^{1/k_a} 
\]
has a minimum, $\stackrel{\vee }{G}$, occurring at 
\[
\stackrel{\vee }{\rho }_d\equiv \left[ \frac{k_dB_0}{k_bD_0}\right] ^{\frac{%
k_ak_d}{k_a+k_d}}\,\,\Leftrightarrow \,\,\stackrel{\vee }{D}=\left[
k_a^{k_d}k_d^{-k_d}B_0^{k_d}D_0^{k_a}\right] ^{1/\left( k_a+k_d\right) } 
\]
Although neither $F$ nor $G$ are functions bounded from above, the
constraint $F+G=1$ does impose upper bounds in this context. Thus, we find
the largest value $F$ can take must be given by $\stackrel{\wedge }{F}=1-%
\stackrel{\vee }{G}$ . Similarly, we have $\stackrel{\wedge }{G}=1-\stackrel{%
\vee }{F}$ . While $\stackrel{\wedge }{F}$ and $\stackrel{\wedge }{G}$ are
unique, each is associated with {\em two} points, in $A$ and $D$ (or $\rho
_a $ and $\rho _d$). Labeling these pairs by $A_{\pm }$ and $D_{\pm }$, we
see their significance: They are the extremal points of the orbit, i.e., 
\[
A\in \left[ A_{-},A_{+}\right] ,\qquad D\in \left[ D_{-},D_{+}\right] ~.
\]
Let us emphasize that $A_{-},D_{-}\neq \stackrel{\vee }{A},\stackrel{\vee }{D%
}$ in general: The former is the lower bound for $A$ while the latter is
associated with the minimum of $F$.

Next, we study the defining equations for the extremal points: 
\[
F\left( A_{\pm }\right) =1-\stackrel{\vee }{G},\qquad G\left( D_{\pm
}\right) =1-\stackrel{\vee }{F}.
\]
These give rise to eqns. (\ref{A-hat eqn}-\ref{KD}) in Section IIIc,
reproduced here for convenience: 
\begin{eqnarray}
A_{\pm }+J_AA_{\pm }^{-k_b/k_a} &=&1-K_A\left( B_0^{k_d}D_0^{k_a}\right)
^{1/\left( k_a+k_d\right) }   \\
D_{\pm }+J_DD_{\pm }^{-k_a/k_d} &=&1-K_D\left( A_0^{k_b}C_0^{k_a}\right)
^{1/\left( k_a+k_b\right) }\,\,.  
\end{eqnarray}
with 
\begin{eqnarray}
J_A &\equiv &C_0A_0^{k_b/k_a},\quad \,J_D\equiv B_0D_0^{k_a/k_d} \\
K_A &\equiv &\left( \frac{k_a}{k_d}\right) ^{k_d/\left( k_a+k_d\right)
}+\left( \frac{k_d}{k_a}\right) ^{k_a/\left( k_a+k_d\right) }   \\
K_D &\equiv &\left( \frac{k_b}{k_a}\right) ^{k_a/\left( k_a+k_b\right)
}+\left( \frac{k_a}{k_b}\right) ^{k_b/\left( k_a+k_b\right) }\,\,.
\end{eqnarray}
Let us reiterate: When a species is at its maximum or its minimum, the
fractions of its opposing partner-pair assume the {\em same value}. Thus,
when $A=A_{\pm }$, $B$ takes on the same value: $\stackrel{\vee }{B}=\left(
k_d^{k_a}k_a^{-k_a}B_0^{k_d}D_0^{k_a}\right) ^{1/\left( k_a+k_d\right) }$
while $D$ assumes the value $\stackrel{\vee }{D}$.

The other pairs of extremal points ($B_{\pm },C_{\pm }$) can be similarly
studied. But, it is much simpler to exploit the invariant $f,g$ and to
recognize that the extremal points are paired, so that 
\[
C_{\pm }=\left( \left. f\right/ A_{\mp }^{k_b}\right) ^{1/k_a},\qquad B_{\pm
}=\left( \left. g\right/ D_{\mp }^{k_a}\right) ^{1/k_d}\,\,. 
\]

Though the solutions to equations like (\ref{A-hat eqn}) should provide some
indications of the extent of the saddle shaped orbits, our goal is the simple
form (\ref{saddle1}) of the orbit equation. We begin with the observation
that both $F\left( \rho_a\right) -\stackrel{\vee }{F}$ and $G\left( 
\rho_d\right) -\stackrel{\vee }{G}$ lie in the range $\left[
0,M\right] $, where 
\[
M\equiv \stackrel{\wedge }{F}-\stackrel{\vee }{F}=1-\stackrel{\vee }{G}-%
\stackrel{\vee }{F}=\stackrel{\wedge }{G}-\stackrel{\vee }{G} ~.
\]
This motivates us to introduce the variables 
\begin{eqnarray*}
\alpha &\equiv &\pm \sqrt{\left. \left[ F\left( \rho_a\right) -%
\stackrel{\vee }{F}\right] \right/ M} \\
\beta &\equiv &\pm \sqrt{\left. \left[ G\left( \rho_d\right) -%
\stackrel{\vee }{G}\right] \right/ M}
\end{eqnarray*}
where the $\pm $ follows the sign of $\rho _m-\stackrel{\vee }{\rho }_m$.
Clearly, $\alpha ,\beta \in \left[ -1,1\right] $ and eqn. (\ref{saddle1}) is
transformed into the familiar one for a {\em unit circle}, 
\begin{equation}
\alpha ^2+\beta ^2=1  ~.\label{unit circle}
\end{equation}
Undoubtedly, there is a global coordinate transformation so that all closed
orbits are just circles of various radii. But, it may be a challenge to find
this transformation and we will not pursue further. Instead, let us note the possibility of
defining an `angle' $\theta $ by 
\begin{equation}
\theta \equiv \tan ^{-1}\sqrt{\frac{G\left( \rho_d\right) -\stackrel{%
\vee }{G}}{F\left( \rho_a\right) -\stackrel{\vee }{F}}}=\tan ^{-1}%
\sqrt{\frac{B+D-\stackrel{\vee }{G}}{A+C-\stackrel{\vee }{F}}} \label{eq:theta}
\end{equation}
with the understanding that appropriate signs of the roots be used for the
four phases. It is clear that $\theta $ is the only variable in our problem,
since we have two invariants and a constraint, see eqns. (\ref{f,g}) and (\ref{sum}%
). In this sense, since the fractions can be written as (complicated,
implicit) functions of $\theta $, the evolution is controlled by a first
order ordinary differential equation $\partial _t\theta ={\cal F}\left(
\theta \right) $. The solution can be formally written as $\int_0^\theta 1/%
{\cal F}=t$ and so the period, $T$, around the closed orbit is $\int_0^{2\pi
}1/{\cal F}$. Beyond such formal considerations, we believe that $\theta $
can be exploited to formulate our problem with action-angle like variables.
Such an approach is likely to play an important role in the study of the
full stochastic model, in which $\theta $ serves as the fast variable, while 
$f,g$ will wander slowly due to the noise \cite{Dob10}.

Let us close this appendix with two remarks. We should emphasize that the
`unit circle' is just associated with a {\em projection} of the saddle-shaped
orbit. Though the time dependence (i.e., motion around the circle) and $T$
remain in general quite implicit, some simplifications do emerge if certain
pairs of rates are equal, e.g., $k_a=k_d$ (and so, $k_b=k_c$). The period,
for example, assumes a less prohibitive form than $\int_0^{2\pi }1/{\cal F}$.
Instead of $\theta $, we define $\phi \equiv \ln \rho _a$ and arrive at 
\[
T=\frac{2k_a}{k_b}\int_{\phi _{-}}^{\phi _{+}}\frac{d\phi }{\sqrt{\left[
1-A_0e^{\phi /k_b}-C_0e^{-\phi /k_a}\right] ^2-4B_0D_0}} 
\]
where the end points are associated with $A_{\pm }$ and are given by 
\[
A_0e^{\phi _{\pm }/k_b}+C_0e^{-\phi _{\pm }/k_a}=1-2\sqrt{B_0D_0}\,\,. 
\]
Of course, this class of rates includes the most special case, with all $k$%
's being equal, in \cite{Sza07}.

\section{Null space of ${\Bbb K}$ and ${\Bbb M}$}

Here, we show that ${\Bbb K}$ has at most two zero eigenvalues, so that the
dimension of its null space is two or less. For odd $S$, the characteristic
equation antisymmetric matrix (${\Bbb K}$) must be of the form 
\begin{equation}
0=\det \left[ {\Bbb K}-\kappa {\Bbb I}\right] =-C_1\kappa +C_3\kappa
^3...-\kappa ^S  \label{det}
\end{equation}
where the $C$'s are the co-factors. It is straightforward to see that $C_1$
is the sum 
\begin{eqnarray*}
&&\det \left( 
\begin{array}{cccccc}
0 & k_2 & 0 & ... & 0 & 0 \\ 
-k_2 & 0 & k_3 & ... & 0 & 0 \\ 
0 & -k_3 & 0 & ... & 0 & 0 \\ 
\vdots & \vdots & \vdots & 0 & \vdots & \vdots \\ 
0 & 0 & 0 & ... & 0 & k_{S-1} \\ 
0 & 0 & 0 & ... & -k_{S-1} & 0
\end{array}
\right) \\
&&+\det \left( 
\begin{array}{cccccc}
0 & 0 & 0 & ... & 0 & -k_S \\ 
0 & 0 & k_3 & ... & 0 & 0 \\ 
0 & -k_3 & 0 & ... & 0 & 0 \\ 
\vdots & \vdots & \vdots & 0 & \vdots & \vdots \\ 
0 & 0 & 0 & ... & 0 & k_{S-1} \\ 
k_S & 0 & 0 & ... & -k_{S-1} & 0
\end{array}
\right) \\
&&+...+\det \left( 
\begin{array}{cccccc}
0 & k_1 & 0 & ... & 0 & 0 \\ 
-k_1 & 0 & k_2 & ... & 0 & 0 \\ 
0 & -k_2 & 0 & ... & 0 & 0 \\ 
\vdots & \vdots & \vdots & 0 & \vdots & \vdots \\ 
0 & 0 & 0 & ... & 0 & k_{S-2} \\ 
0 & 0 & 0 & ... & -k_{S-2} & 0
\end{array}
\right)
\end{eqnarray*}
which is 
\begin{equation}
k_2^2k_4^2...k_{S-1}^2+k_3^2k_5^2...k_S^2+...+k_1^2k_3^2...k_{S-2}^2\,\,.
\label{C1}
\end{equation}
A systematic writing of this expression is 
\[
\sum_{\ell =0}^{S-1}\left[ \prod_{n=1}^{\left( S-1\right) /2}k_{\ell
+2n}^2\right] ~.
\]
Here, indices greater than $S$ are understood to mean $%
%TCIMACRO{\func{mod}}
%BeginExpansion
\mathop{\rm mod}
%EndExpansion
S$, of course. Thus, for example, the product in the $\ell =S-1$ term runs
from $S-1+2\Rightarrow 1$ to $S-1+S-1\Rightarrow S-2$, i.e., the last term
in (\ref{C1}). Since this cofactor is positive, there is just one solution
to eqn. (\ref{det}) with $\kappa =0$. Thus, the null space of ${\Bbb K}$ is
precisely one dimensional (for odd $S$). In the main text, we show that
this result implies the existence of a single (non-trivial) fixed point,
`dual' to a single invariant, see eqns. (\ref{X*},\ref{curly R}).

For even $S$, the rearrangement into `teams' led us to consider ${\Bbb M}$
instead of ${\Bbb K}$, see eqn. (\ref{K-M}). With no particular symmetry, the
characteristic equation for ${\Bbb M}$ is 
\begin{eqnarray*}
0 &=&\prod_{n=1}^s\left( k_{2n-1}-\mu \right) -\prod_{n=1}^sk_{2n} \\
&=&\det {\Bbb M}-\Gamma _1\mu +...
\end{eqnarray*}
with $\Gamma _1$ being $\left( k_1k_3...k_{S-1}\right) \sum_nk_{2n-1}^{-1}>0$%
. Thus, the null space of ${\Bbb M}$ is at most one-dimensional, so that the
null space of ${\Bbb K}$ is either zero- or two-dimensional (for even $S$). 
\begin{eqnarray*}
&& \\
&& \\
&&
\end{eqnarray*}


\newpage

\begin{references}
\bibitem{May76}  R.M. May, % {\em Simple mathematical models with very
% complicated dynamics} 
Nature {\bf 261}, 459 (1976).

\bibitem{Feig78}  M. Feigenbaum, % {\em Quanitative universality for a
% class of nonlinear transformations} 
J. Stat. Phys. {\bf 19}, 25 (1978).

\bibitem{Feig79}  M. Feigenbaum, % {\em The universal metric properties
% of non-linear transformations} 
J. Stat. Phys. {\bf 21}, 669 (1979).

\bibitem{Hof98}  J. Hofbauer and K. Sigmund, {\it Evolutionary Games and
Population Dynamics} (Cambridge University Press, Cambridge, 1998).

\bibitem{Now06}  M. A. Nowak, {\it Evolutionary Dynamics}
(Harvard University Press, Cambridge, 2006).

\bibitem{Sza07}  G. Szab\'{o} and G. F\'{a}th, 2007 {\it Phys. Rep.} {\bf 446}, 97 (2007).

\bibitem{Fre09}  E. Frey, Physica A {\bf 389}, 4265 (2010).

\bibitem{Fra96a}  L. Frachebourg, P. L. Krapivsky, and E. Ben-Naim,
Phys. Rev. Lett. {\bf 77}, 2125 (1996).

\bibitem{Fra96b}  L. Frachebourg, P. L. Krapivsky, and E. Ben-Naim,
Phys. Rev. E {\bf 54}, 6186 (1996).

\bibitem{Pro99}  A. Provata, G. Nicolis, and F. Baras, J. Chem. Phys. {\bf 110}, 8361 (1999).

\bibitem{Tse01}  G. A. Tsekouras and A. Provata, Phys. Rev. E {\bf 65}, 016204 (2001).

\bibitem{Ker02}  B. Kerr, M. A. Riley, M. W. Feldman, and B. J. M. Bohannan,
Nature {\bf 418}, 171 (2002).

\bibitem{Kir04}  B. C. Kirkup and M. A. Riley, Nature {\bf 428}, 412 (2004).

\bibitem{Rei06}  T. Reichenbach, M. Mobilia, and E. Frey, Phys. Rev. E
{\bf 74}, 051907 (2006).

\bibitem{Rei07}  T. Reichenbach, M. Mobilia, and E. Frey, Phys. Rev. Lett.
{\bf 99}, 238105 (2007).

\bibitem{Rei08}  T. Reichenbach, M. Mobilia, and E. Frey, Nature {\bf 448}, 1046 (2008).

\bibitem{Cla08}  J. C. Claussen and A. Traulsen, Phys. Rev. Lett. {\bf 100}, 058104 (2008).

\bibitem{Pel08}  M. Peltom\"{a}ki and M. Alava, Phys. Rev. E {\bf 78}, 031906 (2008).

\bibitem{Rei08a} T. Reichenbach and E. Frey, Phys. Rev. Lett. {\bf 101}, 058192 (2008).

\bibitem{Ber09}  M. Berr, T. Reichenbach, M. Schottenloher, and E. Frey, 
Phys. Rev. Lett. {\bf 102}, 048102 (2009).

\bibitem{Ven10}  S. Venkat and M. Pleimling, Phys. Rev. E {\bf 81}, 021917 (2010).

\bibitem{Shi10}  H. Shi, W.-X. Wang, R. Yang, and T.-C. Lai, Phys. Rev. E
{\bf 81}, 030901(R) (2010).

\bibitem{And10}  B. Andrae, J. Cremer, T. Reichenbach, and E. Frey,
Phys. Rev. Lett. {\bf 104}, 218102 (2010).

\bibitem{Rul10}  S. Rulands, T. Reichenbach, and E. Frey,
arXiv:1005.5704.

\bibitem{Wan10}  W.-X. Wang, Y.-C. Lai, and C. Grebogi, Phys. Rev. E
{\bf 81}, 046113 (2010).

\bibitem{Mob10}  M. Mobilia, J. Theor. Biol. {\bf 264}, 1 (2010).

\bibitem{He10}  Q. He, M. Mobilia, and U. C. T\"{a}uber, Phys. Rev. E {\bf 82}, 051909 (2010).

\bibitem{Win10}  A. A. Winkler, T. Reichenbach, and E. Frey, Phys. Rev. E {\bf 81}, 060901(R) (2010).

\bibitem{Dob10}  A. Dobrinevski and E. Frey, arXiv:1001.5235.

\bibitem{CDPZ-JSTAT} S. O. Case, C. H. Durney, M. Pleimling, and R. K. P. Zia, in preparation.

\bibitem{CDPZ-epl} S. O. Case, C. H. Durney, M. Pleimling, and R. K. P. Zia, EPL {\bf 92}, 58003 (2010).

\bibitem{Fra98}  L. Frachebourg and P. L. Krapivsky, J. Phys. A: Math. Gen. {\bf 31}, L287 (1998).

\bibitem{Kob97}  K. Kobayashi and K. Tainaka, J. Phys. Soc. Jpn. {\bf 66}, 38 (1997).

\bibitem{Sat02}  K. Sato, N. Yoshida, and N. Konno, Appl. Math. Comput. {\bf 126}, 255 (2002).

\bibitem{Sza04}  G. Szab\'{o} and G. A. Sznaider, Phys. Rev. E {\bf 69}, 031911 (2004).

\bibitem{He05}  M. He, Y. Cai, and Z. Wang, Int. J. Mod. Phys. C {\bf 16}, 1861 (2005).

\bibitem{Sza07b}  G. Szab\'{o}, A. Szolnoki, and G. A. Sznaider,
Phys. Rev. E {\bf 76}, 051292 (2007).

\bibitem{Sza08}  G. Szab\'{o} and A. Szolnoki, Phys. Rev. E {\bf 77}, 011906 (2008).

\bibitem{Sil92}  J. Silvertown, S. Holtier, J. Johnson, and P. Dale,
Journal of Ecology {\bf 80}, 527 (1992).

\bibitem{Sza01}  G. Szab\'{o} and T. Cz\'{a}r\'{a}n, Phys. Rev. E {\bf 63}, 061904 (2001).

\bibitem{Sor09} D. Sornette, V. I. Yukalov, E. P. Yukalova, J.-Y. Henry, D. Schwab, and
J. B. Cobb, J. Biol. Syst. {\bf 17}, 225 (2009).

\bibitem{Zia10} R. K. P. Zia, arXiv:1101.0018.

\end{references}
\end{document}